\documentclass[aps,preprintnumbers,superscriptaddress,nofootinbib]{revtex4}
\usepackage{epsfig,graphics}
\usepackage{amsfonts,amssymb,amsmath}            % for math symbols.
\usepackage{pifont}                              % for cross symbol
\usepackage{amsthm}
\usepackage{subfigure}
\def\identity{\leavevmode\hbox{\small1\kern-3.8pt\normalsize1}}

\theoremstyle{plain}

\newcounter{pp}

\begin{document}

\title{A Proposed Test of the Local Causality of Spacetime}
\date{10 July, 2005 (revised July 2008)}

\author{Adrian \surname{Kent}}
\email{a.p.a.kent@damtp.cam.ac.uk}
\affiliation{Centre for Quantum Computation,
             DAMTP,
             Centre for Mathematical Sciences,
             University of Cambridge,
             Wilberforce Road,
             Cambridge CB3 0WA, U.K.}

\begin{abstract}
A theory governing the metric and matter fields in spacetime is {\it locally causal} 
if the probability distribution for the fields in any  
region is determined solely by physical data in the region's 
past, i.e. it is independent of events at space-like separated points.  
General relativity is manifestly locally causal, since the fields
in a region are completely determined by physical data in its past.   
It is natural to ask whether other possible theories in which the fundamental description
of space-time is classical and geometric --- for instance,  
hypothetical theories which stochastically couple a classical 
spacetime geometry to a quantum field theory of matter --- might also
be locally causal. 

A quantum theory of gravity, on the other hand, should 
allow the creation of spacetimes which violate local causality
at the macroscopic level.   
This paper describes an experiment to test the local causality
of spacetime, and hence to test whether or not gravity behaves as 
quantum theories of gravity suggest, in this respect. 
The experiment will either produce direct evidence that the gravitational field 
is not locally causal, and thus weak confirmation of quantum gravity,
or else identify a definite limit to the domain of 
validity of quantum theory. 
\end{abstract}

\maketitle

\section{Introduction}

Abner Shimony's many profound contributions to theoretical physics have 
greatly deepened our understanding of the nature of physical reality.  
This paper is devoted to subjects on which Abner's 
work is particularly celebrated, namely the theoretical
definition and understanding of locality and
local causality and the ways in which these properties
can be experimentally tested in Nature.

General relativity and quantum 
theory are both impressively confirmed within their
domains of validity, but are, of course, mutually
inconsistent.   Despite decades of research,
there are still deep conceptual problems in formulating 
and interpreting quantum gravity theories: we don't have a fully consistent
quantum theory of gravity, nor do we know precisely how we
would make sense of one if we did.  

One initially natural-seeming possibility is combine general relativity and
quantum theory in a semi-classical theory that couples the metric
to the expectation of the stress-energy tensor via
the Einstein equations \cite{moller,rosenfeld, kibble}.  
However, the problems with this suggestion are well-known.  
In particular, if the unitary quantum evolution of the matter fields
is universal, then it would imply that the complete state of
the matter fields in the current cosmological era ought to
be a superposition of many (in fact, presumably an infinite continuum of)
macroscopically distinct cosmologies.
A semi-classical theory of gravity coupled to these matter fields
would imply, inter alia, that the gravitational fields in our solar
system and galaxy correspond to the weighted average over all possible
matter distributions, rather than the actual distribution we observe.
This would be grossly inconsistent with the observed data.
It is also contradicted by terrestrial experiment \cite{pagegeilker}.   

One might try to rescue the hypothesis by supposing, instead, that unitary 
quantum evolution is not universal and that the metric couples to the
expectation of the stress tensor of non-unitarily evolving matter fields.   
Obviously, this requires some 
explicit alternative to unitary quantum theory, such as a dynamical collapse model \cite{grwp}.
It is not presently known whether such a theory can be combined with a metric 
theory of gravity in a generally covariant way.  
An interesting related possibility is that a classical metric might be
coupled to quantum matter via stochastic equations \cite{penrose, diosi}:
however, no consistent and generally covariant theory of this type has yet been developed either. 

I take here a possibly controversial stance.
It seems to me that, because we haven't made any really certain progress in 
understanding how general relativity and quantum theory are unified,
we should take more seriously the possibility that the answer might
take a rather different form from anything we've yet considered.
On this view, even apparently rather basic and solid intuitions are worth questioning: 
if an intuition can be tested experimentally, and we can unearth a sliver of 
motivation for speculating that it might possibly fail, we should test it. 

\section{Gravity, local causality and reality}

\subsection{Sketch of experiment} 

Before getting into technicalities, let me summarise the proposed experiment.  

We start with a standard
Bell experiment, carried out on an entangled pair of elementary
particles, in which the measurement choices and  
measurement outcomes on both wings are spacelike separated.  

The choices and outcomes are then amplified to produce distinct local 
gravitational fields, on both wings.   This amplification can be 
carried out by any practical means, for example by recording the 
choice and outcome on each wing in an electronic signal, and 
feeding this signal into a circuit connected to a device that moves a 
macroscopic quantity of matter to one of four possible macroscopically distinct
configurations.
Note that this amplification need {\it not} necessarily maintain quantum coherence. 

These gravitational fields produced are then directly measured, by observing their
influence on small masses in the relevant region, for example by Cavendish experiments.  
This is done quickly enough that the region $A_{2}$, in which
the amplified gravitational field on wing $A$ is measured, 
is spacelike separated from the region $B_{1}$ in which the 
Bell measurement choice on wing $B$ was made, and similarly $A_1$
is spacelike separated from $B_2$.  The results of these measurements
are recorded and compared, to check whether they display the correlations
which quantum theory predicts for the relevant Bell experiment.  

\subsection{Standard expectations and why they should be tested}

Almost all theoretical physicists  would, I think, fairly confidently predict that
any experiment of this type will indeed produce exactly the same
non-local correlations as those observed in standard Bell
experiments.   What I want to argue is that there are
some coherent -- although of course speculative --
theoretical ideas which would imply a different outcome, and that these provide scientific
motivation enough to justify doing the experiment.  
To justify this, one needn't argue that the standard expectation is likely
to be wrong (indeed, I think it's very likely right). 
One need only argue that there are some alternative lines of thought 
which have some non-negligible probability of being closer to the 
truth.\footnote{Obviously, there's no precise way to quantify how likely
a surprising outcome must be to make an experiment worth doing.
But to give a rough illustration, a probability of $10^{-5}$ of a surprising 
answer here would seem to me more than sufficient justification for carrying
out an experiment that requires relatively modest resources.}

\subsubsection{One possible motivation}

One view of quantum theory, advocated by 
Bell and taken seriously by many, is that the theory is 
incomplete without some mathematical account of ``beables'' or 
``elements of reality'' or ``real events'' --- the 
quantities which, ultimately, define the sample
space for quantum probabilities, i.e. which are the
things which quantum probabilities are probabilities of.  
Most attempts to resolve this problem postulate that the
beables are at least approximately localised in space-time. 

Now, a standard Bell experiment ensures that the particles in
the two wings enter detectors at space-like separated
points, in a sense which we can justify intuitively
within the quantum path integral formalism (and more 
precisely in some interpretations of quantum theory).
But this does not ensure that any beables or real
events associated with the measurements are necessarily
space-like separated.   For instance, if the beables 
or real events are associated with the collapse of
the wavefunction, and if this collapse takes place
only when a measurement result is amplified to
macroscopic degrees of freedom, then the relevant
question is whether these amplification processes
on the two wings take place in space-like separated
regions.     

Consider now:

{\bf Assumption I } \qquad Bell
experiments appear to produce non-local correlations,
consistent with the predictions of quantum theory, 
when the relevant beables are time-like separated 
(i.e. when there is time for information about the
first relevant real event to propagate to the second),
but {\it not} when they are space-like separated.   

{\bf Assumption II} \qquad in all Bell experiments to
date, the relevant beables have indeed been time-like separated.

If both assumptions were correct, the apparent demonstration of non-locality
in Bell experiments to date would be an artefact. 
The assumptions may, however, at first sight seem purely conspiratorial
and completely lacking in theoretical motivation.  
Surprisingly, though, it {\it is} possible to sketch an alternative version 
of quantum theory which appears to be internally consistent, is not 
evidently refuted by the data, and implies both {\bf I} and {\bf II}  \cite{akcausalqt}.  

Now, let us extend this speculation further.   It is sometimes suggested
that the solution to the quantum
measurement problem is tied up with the link between quantum theory
and gravity.  Consider 

{\bf Assumption III} \qquad to ensure that a real event (selecting one outcome and one 
of the possible fields) takes place requires a measurement event whose
different possible outcomes create measurably distinct gravitational 
fields. 

If {\bf (I-III)} were all true, the gravitational Bell
experiment described above would indeed produce a different
outcome from standard Bell experiments. 
To be sure, taking this possibility seriously requires one to take 
seriously three non-standard hypotheses.
From the perspective of a firm believer in the universality of
unitary quantum evolution and in quantum gravity, each of these
hypotheses might be seen as quite implausible.  
It is worth stressing, though, that none of these hypotheses is
an ad hoc invention, produced specifically for the purposes of 
the present discussion.  Each of them has an independent motivation: 

{\bf (I)} results from a nonstandard but interesting way of 
trying to reconcile beable quantum theory and special relativity.   

{\bf (II)} becomes quite plausible if one takes seriously
the idea of wave function collapse as a real physical process
defined by explicit equations.  Models, such as those defined
by Ghirardi-Rimini-Weber-Pearle\cite{grwp}, which have this feature and
which are consistent with other experiments tend to imply that
collapse only takes place quickly (on a scale of $\mu$s) as 
the measurement result becomes amplified to a macroscopic number of
particles (of order $10^{17}$).  In other words, according to these
models, collapse need not take place at all quickly in the 
photo-detectors or electronic circuits
used in standard Bell experiments.  Hence, it need not 
necessarily be the case that there are spacelike separated collapses in 
the two wings of such experiments: as far as I am aware, in all 
Bell experiments to date, reasonable choices of the GRWP
collapse parameters would imply that no significant collapse 
occurs until later, after the data have been brought together and stored.  

{\bf (III)} is a widely considered, if non-standard, intuition
about the possible form of a theory unifying quantum theory
and gravity.  It is also related to another motivation for
the proposed experiment, to which we now turn.  

\subsubsection{A second possible motivation}

Perhaps quantum theory and general relativity are unified, not via
a quantum theory of gravity, but by some theory which somehow combines a 
classical description of a space-time manifold with a metric
together with a quantum description of matter fields.   
Any such theory would presumably have to have a 
probabilistic law for the metric, since it seems essentially impossible
to reconcile a deterministic metric evolution law with 
quantum indeterminism. 
That is, a fundamental law of nature selects a $4$-geometry
drawn from a probability distribution defined by some 
set of principles, which also define the evolution of 
matter.   Also, to be consistent with observation to date, these 
principles must tend to
produce spacetimes approximately described by the Einstein equations
on large scales.

Granted, we don't even know whether there {\it is} a consistent generally
covariant theory of this form.  Before dismissing the 
entire line of thought as thus presently unworthy of attention, though, 
one should remember that we don't know if there's a consistent 
quantum theory of gravity either.   The idea of a stochastic 
hybrid theory, with a classical manifold coupled to quantum
matter, has some attraction, despite its difficulties, as it   
suggests a possible way around some of the conceptual problems that 
arise when trying to make sense of a quantum theory of spacetime.   

Suppose then that we agree to take this idea as serious enough to be worth 
contemplating exploring a little.   
Given the central role of causality in general relativity, 
it seems reasonably natural to consider the class of metric theories 
whose axioms require the metric encode some version of Einstein causality.
Such theories would preclude the gravitational field exhibiting the 
type of non-local correlations that quantum theory predicts
for matter fields --- and so would have surprising and 
counter-intuitive features.   Once again, it needs to be stressed
that we neither want nor need to argue that this is the likeliest possibility,
only that it has {\it some} theoretical 
motivation and has testable consequences.  
In the next section we define a local causality principle adapted to 
non-deterministic metric theories, and examine its consequences. 

\section{Local causality for metric theories: technicalities}

One key feature on which various theories and proto-theories
of gravity differ is the causal structure of the classical or
quasi-classical space-time which emerges.   
Bell's definition of local causality \cite{bellloccaus} applies to
physical operations taking place in a fixed Minkowski space-time.
As Bell famously showed, quantum theory is not locally causal.  
The possibility of adapting the definition to apply to theories 
with a variable space-time geometry (or a variable structure of 
some sort from which space-time geometry is intended to emerge)
has been considered by Rideout and Sorkin \cite{rideoutsorkin} and Henson \cite{henson},
among others.   The following definition is a modified version of 
one suggested by Dowker \cite{dowker}. 

Define a {\it past region} in a metric spacetime to be a region which contains
its own causal past, and the {\it domain of dependence} of
a region $R$ in a spacetime $S$ to be the set of points $p$ such that every
endless past causal curve through $p$ intersects $R$.     

Suppose that we have identified a specified past region of spacetime 
$\Lambda$, with specified metric and matter fields, 
and let $\kappa$ be any fixed region with specified metric and matter fields. 

Let $\Lambda'$ be another past region, again with specified metric and
matter fields.  (In the cases we are most interested in, $\Lambda \cap \Lambda'$ 
will be non-empty, and thus necessarily also a past region.)

Define 
$${\rm Prob}( \kappa | \Lambda \perp \Lambda' )$$ 
to be the probability that the domain of dependence of $\Lambda$ 
will be isometric to $\kappa$, given that $\Lambda \cup \Lambda'$ 
form part of space-time, and given that the domains of dependence
of $\Lambda$ and $\Lambda'$ are space-like separated regions.  

Let $\kappa'$ be another fixed region of spacetime with specified metric
and matter fields.  

Define 
$${\rm Prob}( \kappa | \Lambda \perp \Lambda' ; \kappa' )$$
to be the probability that the domain of dependence of $\Lambda$ 
will be isometric to $\kappa$, given that $\Lambda \cup \Lambda'$ 
form part of space-time, that the domain of dependence of 
$\Lambda'$ is isometric to $\kappa'$, and that the domains of 
dependence of $\Lambda$ and $\Lambda'$ are space-like separated. 

We say a metric theory of space-time is {\it locally causal} if
for all such $\Lambda, \Lambda' , \kappa$ and $\kappa'$ the 
relevant conditional probabilities are
defined by the theory and satisfy
$$
{\rm Prob}( \kappa | \Lambda \perp \Lambda' ) 
=  {\rm Prob}( \kappa | \Lambda \perp \Lambda' ; \kappa' ) \, . 
$$

\section{Testing local causality of metric theories} 

By definition, general relativity is locally causal, since the metric and
matter fields in the domain of dependence $\kappa$ of $\Lambda$ 
are completely determined by those in $\Lambda$ via the Einstein equations
and the equations of motion.  
If we neglect (or believe we can somehow circumvent) the fact that
quantum theory is not locally causal (in Bell's original sense), it would
also seem a natural hypothesis 
that any fundamental stochastic theory 
of space-time, or any fundamental stochastic theory coupling a classical
metric to quantum matter, should be locally causal.
One reason for considering this possibility is that, while it admittedly seems hard to 
see how to frame closed form generally covariant equations for any
theory of this type, it seems particularly hard to see how to 
frame such equations for a non-locally causal theory.
If we allow the evolution of the metric, and hence the causal structure,
at any given point to depend on events at space-like separated points, 
it seems difficult to maintain any notion of causality, or to 
find any other ordering principle which ensures that equations have
a consistent solution.   

However, we should {\it not} expect a quasiclassical space-time emerging from a quantum
theory of gravity to be locally causal, for the following reason.  
Consider a standard Bell experiment carried out on two photons in a
polarization singlet state.  For definiteness, let us 
say that the two possible choices of measurement on either wing are
made by local quantum random number generators, and are 
chosen to produce a maximal violation of 
the CHSH inequality \cite{chsh}.   

We suppose that the two wings of the experiment, $A$ and $B$, are 
fairly widely separated.
Now suppose that the measurement choices and outcomes obtained by the detectors in each wing 
mechanically determine one of four macroscopically distinct configurations. 
To be definite, let us suppose that the Bell experiment is coupled to local Cavendish 
experiments on each wing, in such a way that each of the two settings and two
possible measurement outcomes on any given wing causes one of four
different configurations of lead spheres -- configurations which we 
know would, if the experiment were
performed in isolation, produce one of four macroscopically and testably 
distinct local gravitational fields.   Suppose also that the Cavendish
experiments are arranged so that the local gravitational fields are 
quickly tested, using small masses on a torsional balance in the usual way. 
The separation of the two wings is such that the gravitational field
test on either wing can be completed in a region space-like separated 
from the region in which the photon on the other wing is detected. 

A quantum theory of gravity should predict that the superposition of 
quantum states in the singlet couples to the detectors in either wing
to produce entangled superpositions of detector states, and thence 
entangled superpositions that include the states of the Cavendish experiments, and
finally entangled superpositions of states that include the states of
the local gravitational field.  Extrapolating any of the standard interpretations
of quantum theory to this situation, we should expect to see precisely the same 
joint probabilities for the possible values of the gravitational fields in 
each wing's experiments as we should for the corresponding outcomes in the
original Bell experiment.  As Bell \cite{bellpaps} and Clauser et al. \cite{chsh} showed, 
provided we make the standard and natural (although not logically 
necessary) assumption that the measurement choices in each wing 
are effectively independent from the variables determining the outcome in the other wing, 
these joint probabilities violate local causality in Bell's original sense.  

We now make the further natural assumption that when, as in our proposed
experiment, the measurement choices are made by the outputs of the local
quantum random number generators, the choices made on each wing are 
independent of the metric and matter fields in the past of the 
measurement region on the other wing.  
Then, if $\kappa$ is the region immediately surrounding the measurement choice 
and outcome in one wing of the experiment, $\kappa'$ the corresponding
region for the other wing, $\Lambda$ the past of $\kappa$, and $\Lambda'$
the past of $\kappa'$, we have 
$$
{\rm Prob}( \kappa | \Lambda \perp \Lambda' ) 
\neq {\rm Prob}( \kappa | \Lambda \perp \Lambda' ; \kappa' ) \, . 
$$

Does such an experiment even need to be performed, 
given the impressive experimental confirmation of quantum theory in Bell
experiments to date?   In my view, it does.  

Taking the Bell experiments to date at face value -- that is, 
neglecting any remaining possible loopholes in their interpretation -- they
confirm predictions of quantum theory {\it as a theory of 
matter fields when gravity is negligible}.  
Specifically, they confirm predictions of quantum theory for experiments involving
matter states when those states do not produce significant superpositions of 
macroscopically distinct gravitational fields.   

The question at issue
here is precisely how far quantum theory's domain of validity extends.  
When it comes to predicting whether or not the metric is locally causal,
there is a genuine tension between intuitions extrapolated from quantum theory 
and those which one might extrapolate from general relativity.  
Examining and testing this question seems 
a very natural development of the line of questioning 
begun by Einstein, Podolsky and Rosen \cite{epr} and continued by Bell \cite{bellpaps}.  

Standard Bell experiments test the conflicting predictions implied by quantum theory
and by EPR's intuitions about the properties of elements of
physical reality.  EPR's intuitions can be motivated by a combination of 
classical mechanics (which suggests that the notion of an element of 
physical reality is a sensible one) and special relativity (which
suggests the hypothesis that an element of physical reality has 
the locality properties ascribed to it by EPR).  
In the experiment considered here, we again have a tension 
between intuitions drawn from two successful theories -- in this
case quantum theory and general relativity.

\section{Possible Counterarguments}

But isn't this a crazy line of thought?   How could the correlations obtained from Bell 
experiments {\it possibly} be altered by coupling classical devices to the detector outputs?    
Is the Bell experiment supposed to know that the classical devices are waiting for the
data, and change its result because of that?   
Or, even more weirdly, is the gravitational field in each wing supposed to know that the classical 
lumps of matter are being moved around as the result of a Bell experiment,
and change {\it its} behaviour --- violating the predictions of Newtonian
gravity as well as general relativity within a local region --- because 
of {\it that}?  

I find it hard to accept the full rhetorical force of such objections, 
natural though they are. 
Nature has a capacity to surprise, and surprising experimental results 
sometimes have theoretical explanations which occurred to nobody beforehand.   
The ``common sense'' view just expressed implicitly assumes, among other things,
first, that the outcomes of detector measurements in Bell 
experiments constitute local, macroscopic events that in 
some physically meaningful sense are definite and irreversible once they occur, and second, 
that the local gravitational fields respond instantly to these events in the same way
as they would if they resulted from isolated experiments
on unentangled states.   
These plausible propositions may very well be given precise meaning 
and completely justified by some deeper understanding of quantum theory and gravity than we
currently have.  Even if they don't turn out to have a precise and literal justification
--- for instance, because the fundamental theory contains no definition of definite 
local events --- it seems very plausible that we nonetheless reach the right conclusion 
about Bell experiments and gravity by reasoning as though they were true.  
However, none of this is completely beyond reasonable doubt in the light of our 
current knowledge. 

As we've already noted, there's some independent
motivation for exploring variants of quantum theory in which definite local
events {\it are} defined but in which photo-detector measurement outcomes
aren't, so to speak, macroscopic enough to constitute such events. 

There's also some motivation for exploring theories of quantum theory
and gravity in which a probabilistic law defines a locally causal classical 
gravitational field.  Standard reductionist reasoning would break down in
such a theory --- as it does, though in a different way, in quantum theory --- 
and the behaviour of the gravitational field in one wing of a Bell experiment
would indeed depend on the configurations of both wings of the experiment.   

What, then, are the conceivable experimental outcomes, and what would
they imply?  
One is that the violations of local causality predicted by
quantum theory, and to be expected if some quantum theory 
of gravity holds true, are indeed observed.  This would 
demonstrate that space-time is indeed not locally causal,
as predicted by quantum theories of gravity, but not
necessarily by other hypotheses about the unification 
of quantum theory and gravity.
It would thus provide at least some slight experimental evidence in favour
of the quantization of the gravitational field. 
It might be argued, pace Page and Geilker \cite{pagegeilker}, that this 
would be the first such experimental evidence, since, as noted above, Page and
Geilker's experiment tested a version of semi-classical gravity already
excluded by astronomical and cosmological observation.   

A second logical possibility is that the violations of local causality predicted by quantum 
theory fail to be observed at all in this particular extension of 
the Bell experiment: i.e., that the measurement results obtained
from the detectors fail to violate the CHSH inequality.  
This would imply that quantum theory fails to describe correctly
the results of the Bell experiment embedded within this particular
experimental configuration, and so would imply a definite limit
to the domain of validity of quantum theory.

A third logical possibility is that the Bell experiment correlations follow the 
predictions of quantum theory, but that the Cavendish experiments
show gravitational fields which do not 
correspond to the test mass configurations in the expected 
way (or at least do not do so until a signal has had time to travel from one
wing to the other).   This would suggest the coexistence of
a quantum theory of matter with some classical theory of
gravity which respects local causality, but which has the 
surprising property that classical gravitational fields do
not always couple to macroscopic matter in the way suggested by general 
relativity.    

In summary: although our present understanding of physics leads us to expect
the first outcome, the point at issue seems sufficiently fundamental, and 
our present understanding of gravity sufficiently limited, that
it would be very interesting and worthwhile to carry out experiments
capable of discriminating between some (and of course, ideally, all) 
of the possible outcomes outlined above.  
    
\acknowledgements
I am particularly indebted to Fay Dowker for very helpful comments
on an earlier draft, for inspiring the definition
of local causality used in the present version of the paper, and
for many thoughtful criticisms.
I am also very grateful to Nicolas Gisin, Valerio Scarani, Christoph Simon and 
Gregor Weihs for valuable discussions on various criteria for gravitationally
induced collapse and experimental tests.  
Warm thanks too to Harvey Brown, Jeremy Butterfield, Robert Helling, Graeme Mitchison,
Roger Penrose and Rainer Plaga for some very helpful comments.
Last but by no means least, I would very much like to thank Abner for his 
characteristically kind and warm encouragement to pursue this idea.

\end{document}